\documentclass[12pt]{article}
\usepackage{times}
\usepackage{geometry}
\usepackage[utf8]{inputenc}
\usepackage{enumitem,amssymb}
\usepackage{ragged2e}
\usepackage{hyperref}
\newlist{thematic}{itemize}{8}
\setlist[thematic]{label=$\square$}
\usepackage{pifont}
\usepackage[round]{natbib}
\usepackage{url}
\usepackage{graphicx}	
\newcommand{\xmark}{\ding{55}}%

\newcommand{\crossbox}{\rlap{$\square$}{\large\hspace{1pt}\xmark\ }}

\newcommand{\ANTARES}{{\textsc {antares}}}

\newcommand{\apj}{{ApJ}}
\newcommand{\apjl}{{ApJL}}     
\newcommand\pasp{{PASP}}

\newenvironment{my_enumerate}{
\begin{enumerate}
  \setlength{\itemsep}{1pt}
  \setlength{\parskip}{0pt}
  \setlength{\parsep}{0pt}}{\end{enumerate}
}

\newenvironment{my_itemize}{
\begin{itemize}
  \setlength{\itemsep}{1pt}
  \setlength{\parskip}{0pt}
  \setlength{\parsep}{0pt}}{\end{itemize}
}

\begin{document}
\raggedright
\huge
Astro2020 Activities, Projects, or State of the Profession Consideration White Paper \linebreak

Infrastructure and Strategies for Time Domain and MMA and Follow-Up \linebreak
\normalsize

\noindent \textbf{Type of Activity:} \hspace{0.5cm}
\crossbox Ground Based Project \hspace{0.5cm} 
$\square$ Space Based Project \linebreak 
\crossbox Infrastructure Activity \hspace{0.5cm}
\crossbox Technological Development Activity \linebreak
$\square$ State of the Profession Consideration \hspace{0.5cm}
$\square$ Other \linebreak
  
\textbf{Principal Author:}

Name:	Bryan W. Miller
 \linebreak						
Institution:  Gemini Observatory
 \linebreak
Email: bmiller@gemini.edu
 \linebreak
Phone:  +56 51 2205618

\parskip 1.5ex 

\textbf{Co-authors:} 
Lori Allen (NOAO), Eric Bellm (LSST/University of Washington), Federica Bianco (University of Delaware), John Blakeslee (Gemini), Robert Blum (LSST), Adam Bolton (NOAO), Cesar Brice{\~n}o (NOAO/SOAR), Will Clarkson (University of Michigan, Dearborn), Jay Elias (SOAR), Suvi Gezari (University of Maryland), Bob Goodrich (GMT), Matthew Graham (ZTF/Caltech), Melissa Graham (LSST/University of Washington), Steve Heathcote (NOAO), Henry Hsieh (Planetary Science Institute), Jennifer Lotz (Gemini), Tom Matheson (NOAO), M. Virginia McSwain (Lehigh University), Dara Norman (NOAO), Travis Rector (University of Alaska), Reed Riddle (Caltech/ZTF), Stephen Ridgway (NOAO), Abi Saha (NOAO), Rachel Street (Las Cumbres Observatory), Marcelle Soares-Santos (Brandeis University), Warren Skidmore (TMT), Letizia Stanghellini (NOAO), Lou Strolger (STScI), Joanna Thomas-Osip (Gemini), and Kathy Vivas (NOAO) 


\textbf{Abstract:}
Time domain and multi-messenger astrophysics are growing and important modes of observational astronomy that will help define astrophysics in the 2020s. Significant effort is being put into developing the components of a follow-up system for dynamically turning survey alerts into data. This system consists of: 1) brokers that will aggregate, classify, and filter alerts; 2) Target Observation Managers (TOMs) for prioritizing targets and managing observations and data; and 3) observatory interfaces, schedulers, and facilities along with data reduction software and science archives. These efforts need continued community support and funding in order to complete and maintain them. Many of the efforts can be community open-source software projects but they will benefit from the leadership of professional software developers. The coordination should be done by institutions that are involved in the follow-up system such as the national observatories (e.g.\/ LSST/Gemini/NOAO Mid-scale/Community Science and Data Center) or a new MMA institute. These tools will help the community to produce the most science from new facilities and will provide new capabilities for all users of the facilities that adopt them. 

\pagebreak
\textbf{\large{1 Key Science Goals and Objectives}}

\textbf{1.1 Introduction}

The Large Synoptic Survey Telescope \citep[LSST;][]{ivezic2008ab} is currently under construction on Cerro Pach\'{o}n near La Serena, Chile, and, based on experience with the high rate of transient candidates from the Zwicky Transient Facility \citep[ZTF;][]{bellm2019aa}, will take time-domain astronomy to the next level. LSST will image the entire visible sky every few nights for ten years in order to identify variable and transient objects and generate deep images. LSST will produce thousands of alerts every few minutes from variable stars, AGN, solar system bodies, all varieties of cosmic explosions, and transients of as yet unknown forms \citep{ridgway2014aa}. It will not be possible to understand the nature of many of these detections from LSST photometry alone, so additional observations will be required, often on very short timescales after the initial discovery. 

Astronomy entered a new era of multi-messenger astrophysics (MMA) in 2017. The Advanced LIGO and Virgo detectors observed a gravitational wave signal consistent with a neutron star - neutron star merger and a massive observing campaign by electromagnetic (gamma rays to radio) and neutrino telescopes detected the remains of the resulting ``kilonova'' explosion \citep{abbott2017aa}. A month later a muon produced by a collision with a 290 TeV neutrino was detected by the IceCube Neutrino Observatory. The neutrino's origin was coincident with a gamma-ray blazar and follow-up observations from gamma-ray to radio frequencies showed that the blazar was in a ``flaring'' state and that blazars can be a source of high-energy neutrinos \citep{icecube2018aa}. A month after that the Pan-STARRS1 survey detected a fast-moving object on a hyperbolic orbit. Follow-up observations with a large number of optical and IR telescopes showed that `Oumuamua was the first detected body from another solar system \citep{meech2017ab}. These examples show that combining information from different messengers (gravitational waves, neutrinos, high-energy particles, and electromagnetic radiation) collected on short timescales allows us to understand the nature of the sources and the physical processes involved. In the era of TDA and MMA, timely, often rapid follow-up with a multiplicity of flexibly-scheduled observing facilities is essential, and the demand is growing rapidly. 

MMA is also one of the National Science Foundation's (NSF) Ten Big Ideas. It unifies the capabilities and results from some of the NSF's major astrophysics facilities (e.g.\ LIGO, IceCube, LSST, Gemini, KPNO/CTIO, and SOAR) to discover and characterize new and rare events.  The NSF-sponsored report by the National Research Council on \emph{Optimizing the U.S.\ Ground-Based Optical and Infrared Astronomy System} \citep{elmegreen2015aa} recommends that NSF facilities coordinate to optimize LSST follow-up. The same capabilities are needed for MMA follow-up.
 
The sheer volume of alerts means that existing methods which rely on human review for filtering and responding to alert streams will not be able to keep up with LSST and many interesting targets could be missed unless new technologies and methods are developed. The most efficient use of scarce astronomer and telescope time will be attained if access to all available facilities can be streamlined, especially since many science goals will require access to multiple facilities (aperture sizes, instruments, and  wavelengths). Therefore, it is imperative that the community develop systems that can handle the volume of LSST alerts and organize the use of the available follow-up resources. This white paper will summarize the current state of TDA/MMA follow-up preparation and identify some of the important needs for the next decade.

\textbf{1.2 Related Astro2020 White Papers}

Over 40 Astro2020 Science white Papers relevant to MMA and TDA science have been submitted. Many specifically mention the infrastructure and issues discussed here or will be able to take advantage of the system that is described. The number of science cases shows the importance of this initiative. The papers include:

\begin{my_enumerate}
\item Beaton, R.~L., et al., Measuring the Hubble Constant Near and
    Far in the Era of ELTs
\item Blakeslee, J., et al., Probing the Time Domain with High
    Spatial Resolution
\item Brown, P., et al., Keeping an Ultraviolet Eye on Supernovae
\item Burns, E., et al., A Summary of Multimessenger Science with
    Neutron Star Mergers 
\item Burns, E., et al., Opportunities for Multimessenger Astronomy
    in the 2020s
\item Caldwell, R., et al., Cosmology with a Space-Based
      Gravitational Wave Observatory
\item Chang, P., et al., Cyberinfrastructure Requirements to Enhance
  Multi-messenger Astrophysics
\item Chanover, N., et al., Triggered High-Priority Observations of Dynamic Solar System Phenomena
\item Chornock, R., et al., Multi-Messenger Astronomy with
        Extremely Large Telescopes 
\item Cowperthwaite, P., et al., Joint Gravitational Wave and
  Electromagnetic Astronomy with LIGO and LSST in the 2020s
\item Foley, R., et al., Gravity and Light: Combining Gravitational
    Wave and Electromagnetic Observations in the 2020s
\item Graham, M., et al., Discovery Frontiers of Explosive Transients: An ELT and LSST Perspective
\item Hlo\v{z}ek, R., et al., Single-Object Imaging and Spectroscopy
    to Enhance Dark Energy Science from LSST
\item Kara, E., et al., X-ray Follow Up of Extragalactic Transients
\item Kasliwal, M., et al., The Dynamic Infrared Sky
\item Kim, A., et al., Testing Gravity Using Type Ia Supernovae
      Discovered by Next-Generation Wide-Field Imaging Surveys
\item Kupfer, T., et al., A Summary of Multimessenger Science with
    Galactic Binaries
\item Littenberg, T., et al., Gravitational Wave Survey of Galactic Ultra Compact Binaries
\item Lu, J., et al., From Stars to Compact Objects: The Initial-Final Mass Relation
\item Mandelbaum R., et al., Wide-field Multi-object Spectroscopy to Enhance Dark Energy Science from LSST
\item Maccarone, T., et al., Populations of Black holes in Binaries
\item Metzger, B., et al., Kilonovae: nUV/Optical/IR Counterparts of Neutron Star Binary Mergers with TSO
\item Milisavljevic, D., et al., Achieving Transformative Understanding of Extreme Stellar Explosions with ELT-enabled Late-time Spectroscopy
\item Olsen, K., et al., Science Platforms for Resolved Stellar Populations in the Next Decade
\item Newman, J., et al., Deep Multi-object Spectroscopy to Enhance Dark Energy Science from LSST
\item Pasham, D., et al., Probing the Cosmological Evolution of Super-massive Black Holes using Tidal Disruption Flares
\item Perlmutter, S., et al., The Key Role of  Supernova
      Spectrophotometry in the  Next-Decade Dark Energy Science
      Program
\item Rix, H.-W., et al., Binaries Matter Everywhere: from Precision
  Calibrations to Re-Ionization and Gravitational Waves
\item Reitze, D., et al., The US Program in Ground-Based
    Gravitational Wave Science: Contribution from the LIGO Laboratory
\item Ross, N., et al., Opportunities in Time-Domain Extragalactic
    Astrophysics with the NASA Near-Earth Object Camera (NEOCam)
\item Sathyaprakash, B., et al., Multimessenger Universe
      with Gravitational Waves from Binary Systems
\item Scolnic, D., et al., The Next Generation of Cosmological
   Measurements with Type Ia Supernovae
\item Schaefer, G., et al., High Angular Resolution Astrophysics in
   the Era of Time Domain Surveys
\item Shawhan, P., et al., Multi-Messenger Astrophysics
        Opportunities with Stellar-Mass Binary Black Hole Mergers
\item Shoemaker, D., et al., Gravitational-Wave Astronomy in the
      2020s and Beyond: A View across the Gravitational Wave Spectrum 
\item Shoemaker, D., et al., Gravitational Wave Astronomy with LIGO
  and Similar Detectors in the Next Decade
\item Siemiginowska, A., et al., The Next Decade of
      Astroinformatics and Astrostatistics
\item Slosar, A., et al., Dark Energy and Modified Gravity
\item Wang, L., et al., JWST: Probing the Epoch of Reionization
      with  a Wide Field Time-Domain Survey
\item Wheeler, J.~C., et al., ELT Contributions to Tidal Disruption
     Events
\item Wheeler, J.~C., et al., ELT Contributions to The First Explosions
\item Wood-Vasey, M., et al., Type Ia Supernova Cosmology with TSO
\item Zemcov, M., et al., Opportunities for Astrophysical
        Science from the Inner and Outer Solar System
\item Zingale, M., et al., MMA SAG: Thermonuclear Supernovae
\end{my_enumerate}

Other related Activity, Projects, or State of the Profession Consideration (APC) White Papers that are related to this paper or that expand on various aspects of the system discussed here include:
\begin{my_itemize}
    \item Allan, L., et al., The NOAO Mid-Scale Observatories
    \item Bellm, E., et al., Scheduling Discovery in the 2020s
    \item Bolton, A., et al., Community Science and Data-Intensive Astronomy Support at the US National Optical Astronomy Observatory
    \item Matheson, T., et al., \ANTARES: Enabling Time-Domain Discovery in the 2020s
    \item O'Meara, J., et al., The need for robust, near real-time data services on large, ground-based OIR facilities
    \item Levi, et al., The Dark Energy Spectroscopic Instrument (DESI)
    \item Sivo, G., et al., Entering into the Wide Field Adaptive Optics Era in the Northern Hemisphere
    \item Tollerud, E., et al., Sustaining Community-Driven Software for Astronomy in the 2020s
\end{my_itemize}

\textbf{\large{2 Technical Overview: A Follow-up Network}}\label{sec:tech_overview} 

Following the \cite{elmegreen2015aa} recommendation, NOAO and LSST with the support of the Kavli Foundation organized a workshop and report on \emph{Maximizing the Science in the Era of LSST} \citep{najita2016aa} which described the science cases and requirements for LSST follow-up and the basic concept of a follow-up system. More details of an alert follow-up systems were discussed at the May 2017 NOAO workshop ``Building the Infrastructure for Time-Domain Alert Science in the LSST Era'' and are given in the online presentations\footnote{\url{https://www.noao.edu/meetings/lsst-tds/}}. 

\begin{figure*}[t]
    \centering
    \includegraphics[width=\columnwidth]{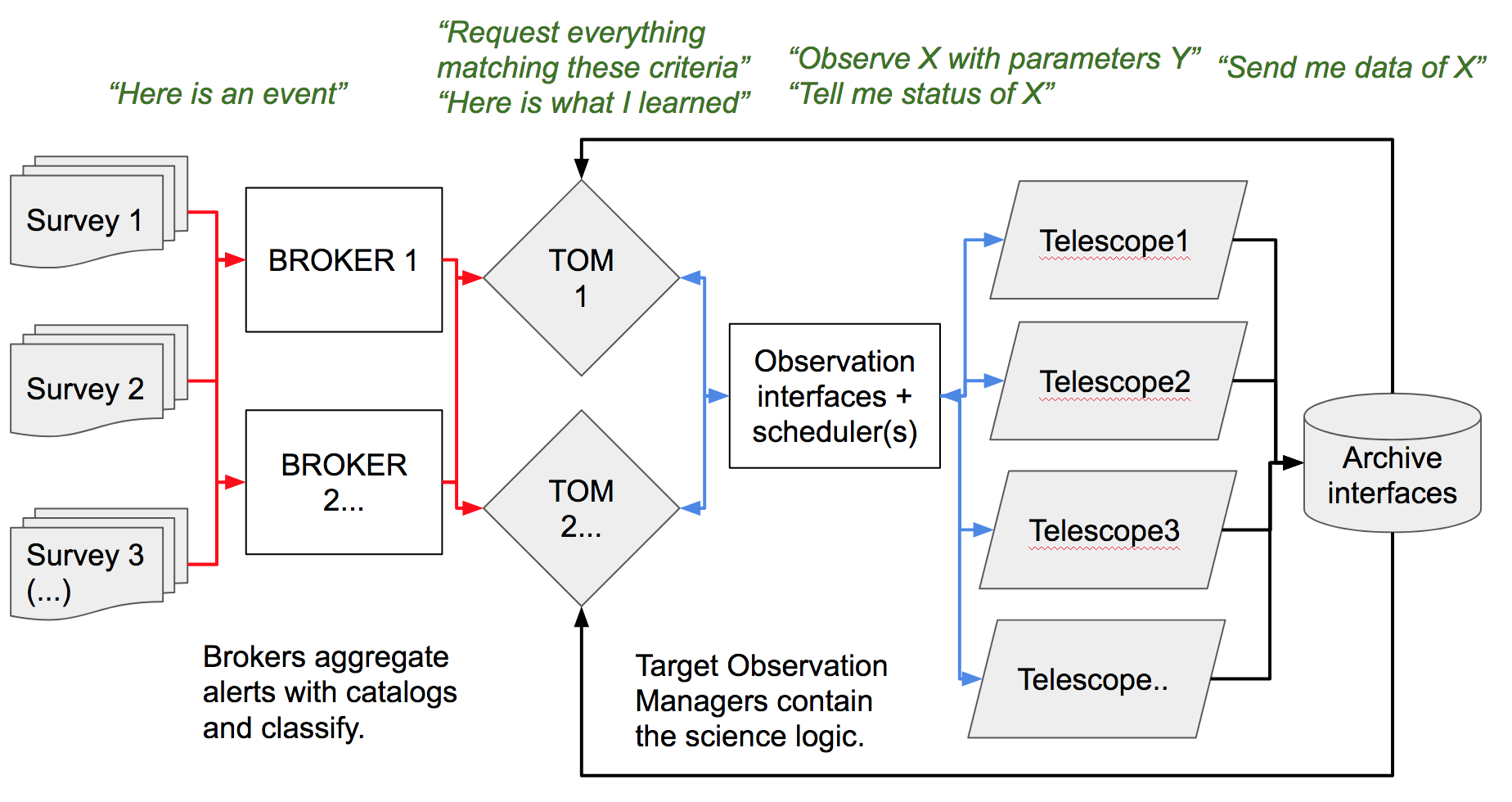}
    \caption{A network flow diagram for transient follow-up from alert streams. Brokers classify alerts. Target Observation Managers (TOMs) then match the targets of interest with the facilities on which science teams have observing time. Finally, systems of telescopes must be available to receive and schedule observations and provide data products via archives (Figure concept: Rachel Street, Las Cumbres Observatory).}
    \label{fig:too_network}
\end{figure*}

The main components of the system from these workshops for dynamically turning alert streams into data are shown in Figure~\ref{fig:too_network}. This system consists of: 1) brokers that will aggregate, classify, and filter alerts; 2) Target Observation Managers (TOMs) that are used by science teams for prioritizing targets and managing observations and data; and 3) observatory interfaces, schedulers, and facilities along with data reduction software and science archives. 
 
The first component is an alert ``broker'' or ``aggregator'' that collects alerts from transient surveys (e.g.\ ZTF, LSST) or facilities (e.g. LIGO via GCN circulars). Events are classified based on cross-matches with existing catalogs and with photometric and astrometric properties (colors, brightness changes, light curves, positions, proximity to other objects). These services will make use of machine-learning and other advanced statistical methods (the technology drivers listed in \S~3). Brokers typically provide sub-streams of alerts filtered by science-driven criteria, made available to users or software agents via, for example, web browser interfaces, application programming interfaces (APIs), and/or query-and-download functionality. The characterized events can then be filtered for objects of interest and the results sent to new alert streams and databases. 

After the alert broker a science layer is needed so that research teams can match the events of interest from the brokers with telescope resources ({\it in real time}). These TOMs present the known information about the objects of interest and aid the science teams with prioritization, managing telescope allocations and observations, collecting and even processing of data, and information access \citep{street2018aa}. A number of TOM systems have been independently created by groups studying supernovae, microlensing, AGN, and solar system objects. 

Currently the communication of ``target-of-opportunity'' (ToO) observations that need rapid reaction times includes phone calls to observers, web-page lists of targets, and electronic ``messages'' or programmatic observation submissions to observatory systems. For maximum efficiency and speed it is expected that the TOMs will send observation requests to the observatories programmatically via application programming interfaces (APIs) in real time, utilizing resources that are available at that time or later as appropriate (i.e., observations are scheduled dynamically in this framework). These APIs should provide the current observing status of the facilities (e.g.\ closed, open, ToO observation status), the available instruments, the ability to receive observation requests, and observation status (e.g.\ pending, observed). 

The observatories then need to execute them efficiently and on the appropriate timescale. Since very rare or rapidly evolving events may need to be observed quickly, dynamical scheduling that can change the observing plan in real time and without significant human interaction is required. Queue-based observatories such as Las Cumbres and Gemini will schedule regular queue observations whenever not responding to rapid triggers. Most rapid-response ToO observations also require longer-term observations to track the evolution of the target and most observations of variable sources require observations during specific periods of time even if they are not interrupting. Regular queue observations often have conditions constraints. Dynamical scheduling allows all of these complex constraints to be met efficiently and fairly.

Some projects such as the study of AGN or X-ray binaries require simultaneous observations using multiple facilities in different wavelength regimes. Currently this is often done via email or telephone. This is inefficient since it is difficult to determine when all the facilities are available. Therefore, it would be very useful to develop a system with which facilities can coordinate observations programmatically. 

In time domain astronomy rapid scientific feedback from observed targets helps to optimize the use of oversubscribed follow-up resources and can enable new science. The overall strategy for coordinating follow-up networks may change with the addition of automated pipeline reductions that close the loop with schedulers. For example, consider a transient with rapidly evolving spectral emission lines but a relatively constant photometric color: the prioritization of such a transient for additional spectroscopic follow-up hinges on the rapid (automated) reduction and analysis of initial follow-up. Automated reduction and analysis would allow for the machine classification of such targets of interest and for the automated submission for further monitoring and/or to invoke changes in the (available) instrument configurations. Large follow-up surveys of static sources will also benefit from automated reduction pipelines. 

At a minimum the raw data needs to be made available immediately after they are taken in a way so that they can be downloaded programmatically. The science teams would then be responsible for reducing the data. This analysis could be conducted by, or in conjunction with, a TOM system, and the results automatically ingested and used by the TOM in planning future observations. It would be desirable for the observatories to provide reduced data products so that the data can be processed using the best practices for that facility and so that each science team is not required to duplicate data reduction pipelines.  The time needed to process a dataset depends on the mode and the length of the observation, but a goal should be to provide reduced longslit data within 30 minutes of it being taken. 

\textbf{Supporting Observing Modes for Networked Follow-Up} 

Although the time-sensitive nature of transient source follow-up prompted the demand for the system described in \S~2, it is important to note that the tools and capabilities will be valuable for all science projects. Many experiments will select targets from survey catalogs or broker databases. TOMs are useful for managing any large observing project. Dynamic scheduling and data reduction pipelines benefit all programs in the queue. The efficiency of a global scheduler or coordinating service will benefit large survey programs and/or multi-wavelength observing campaigns for static sources, and it will reduce the chance of duplicate observations from multiple facilities.

Joining facilities with different observing styles, policies, and allocation processes into a common network for follow-up observations is a significant challenge. While common communication interfaces (e.g. software APIs) are required and must be documented, this is relatively straightforward. The larger challenge is negotiating the politics and sociology of the different organizations, especially because the envisioned follow-up network will be a new observing paradigm for many observers.

The facilities that will join the AEON network described below currently operate in very different ways, and these differences will have to be resolved to move forward. For example, SOAR and the CTIO Blanco 4m have traditionally been scheduled projects in fixed blocks of time. Project teams are then responsible for taking the data themselves. ToOs are handled somewhat ``manually'' with the aid of the current observers or remote ToO teams\footnote{\url{https://www.noao.edu/noaoprop/help/too.html}}. Compensation to interrupted programs can be difficult to reschedule. On the other hand, Gemini and Las Cumbres already operate mostly or exclusively in queue mode in which observations are executed by the observatories based on science rankings, visibility, and various observing constraints (times, conditions, etc). ToOs are handled naturally since observing plans can be updated quickly and any interrupted observations just return to the queue to be scheduled again. Therefore, facilities must offer flexible time for TDA science and clearly document their policies.


\textbf{Modifying Time Allocation Policies for ToO Follow-Up}

Many projects also need access to multiple facilities. For example, the study of the explosions resulting from binary neutron-star mergers detected with LIGO requires imaging surveys followed by spectroscopic characterization.  Supernovae studies may need different aperture telescopes depending on the brightness of the source, or they may use smaller telescopes for classification followed by larger facilities for higher-resolution observations. Therefore, it would be more efficient if investigators could request time on multiple facilities with a single proposal. 

Additionally, proposals for ToOs may require modifications to allow proposers to specify what would constitute a ``duplication'' of their observations in terms of the instrumental configuration and the time-frame in which data is acquired. Observatories may need this information to make on-the-fly decisions in the case of multiple programs requesting similar observations of a high priority target (see the next section). Proposals might also require modifications to allow science-driven requests for exemptions from rapid ToO interrupts, e.g., if the observations are time sensitive. Future Time Allocation Committees would have to be prepared to validate these requests as part of the evaluation process. Finally, new metrics that should be developed to evaluate surveys (e.\/g.\/ urgency vs.\/ importance in scheduling) and priorities of observations so that ToO and non-ToO proposals can be evaluated using similar criteria. 

\textbf{Reducing Redundancy in Follow-Up Observations}

Follow-up resources will be limited in the LSST era, unable to cover the total number of scientifically valuable targets, and it is in the astronomical community?s best interest to develop policies that reduce (or better, eliminate) targets being observed multiple times. It is conceivable that multiple parties may have similar accepted programs to follow-up sources with a given facility, and request to follow-up the same target. Whether or not the community undertakes a major effort to share such data, individual telescope facilities will likely want to enact policies that define and govern the cases in which requested observations constitute a duplication, and in which scenarios the telescope might refuse to take redundant observations. 

For example, in order to reduce redundancy, Gemini Observatory's Policies for Competitive ToOs states that {\it ``ToO data taken for a given PI will be shared with PIs of other programs who also trigger on the same target and ask for the same instrumental configuration''} \citep{GeminiToOPolicy}. The targeted object's evolutionary timescales are taken into consideration when accessing what constitutes a ``duplication''. 

New infrastructure could enable productive cooperation among the user community for astronomers that create TOMs by allowing people to share their follow-up schedules or announce the metadata for completed or planned observations, and possibly even share their non-proprietary data. These tools should take advantage of proposed virtual observatory (VO) standards for target visibility and facility schedules \citep{ness2019aa}.

\textbf{\large{3 Technology Drivers}}\label{sec:tech_drivers}

Technology development is needed in the following areas:
\begin{my_itemize}
    \item Machine learning and other statistical methods for classifying alerts in the brokers 
    \item Algorithms, tools, and communication processes by science teams for prioritizing objects for follow-up
    \item TOM systems and related science-specific analysis tools and interfaces
    \item Algorithms and systems for coordinating observations between different facilities.
    \item Algorithms and systems for rapid dynamic scheduling. Many scheduler solvers exist with different performances and costs. While different approaches may be needed in different situations, at the moment each facility develops their own system from scratch. A common library or toolkit for observation scheduling would make it easier for a facility to implement the capabilities needed to participate in the follow-up system. This would include development for scheduling highly multiplexed spectroscopic facilities (e.g. DESI).
    \item Common data reduction routines (preferably in Python) and pipeline environments.
    \item Archives, algorithms, and systems for rapidly incorporating feedback/results from initial follow-up observations into observation managers to aid in prioritizing further observations.
\end{my_itemize}

\textbf{\large{4 Organization, Partnerships, and Current Status}}

Several groups are developing follow-up systems that are organized similar to that in Figure~\ref{fig:too_network}. Some groups sign memoranda of understanding (MoUs) with different facilities both for receiving alerts and following-up. For example, the AMON system analyzes alerts from gamma ray, X-ray, and neutrino facilities and then sends triggers to a variety of additional facilities for follow up \citep{ayala2019aa}. Their software platform acts like both a broker and a TOM.  Other follow-up networks include PROMPT/SkyNet \citep{reichart2005}, ASAS-SN \citep{kochanek2017aa}, and the Catalina Sky Survey\footnote{\url{http://www.lpl.arizona.edu/css/}}. Additional groups are developing individual components of the system using common standards so that the pieces can work together \citep[e.g.][]{saha2016aa}. 

Many research groups in the USA and worldwide are developing brokers to ingest and process alerts from time-domain surveys (e.g., optical imaging from the ZTF), enabling TDA now and preparing the groundwork for LSST. The LSST will generate $\sim10$ million alerts per night, and distribute the full alert stream directly to at least five brokers (conservatively limited by bandwidth estimates at this time), to be selected via a proposal process which is currently underway \citep{LDM-612}. Representatives of broker teams who submitted Letters of Intent to receive LSST alerts, the first stage of this process, recently met with LSST Project staff at the LSST Community Broker Workshop\footnote{\url{https://ls.st/cbw}}. One of the main topics of discussion was how to collaborate on common infrastructure and services, such as are discussed in this white paper, which could enable LSST alerts to reach as wide a user base as possible. 

Several Target Observation Managers (TOMs) are currently in use by teams studying supernovae, exoplanets, near-earth objects, AGN, and microlensing events \citep{street2018aa}. Until now each team has had to develop its own custom software, at great effort and expense. Fortunately, some of the astronomers at Las Cumbres Observatory have experience with the early TOMs and have identified common required functionality. These features have now been made part of the TOM Toolkit\footnote{\url{https://tomtoolkit.github.io}} that will make it easier for any team to deploy a TOM. Gemini has developed a plugin for the TOM Toolkit so that ToOs can be triggered using the existing, limited API. A video showing some of the TOM's functionality and triggering Gemini observations can be found on YouTube\footnote{\url{https://youtu.be/PC\_5kmSdZBU}}. The TOM Toolkit is now being used to follow-up LIGO events on Las Cumbres and Gemini. A workshop and observing program is also being organized by Las Cumbres Observatory, NOAO, SOAR, and Gemini to promote the development and use of the TOM Toolkit and associated telescopes\footnote{\url{https://lco.global/workshops/tom-toolkit-community-workshop/}}. This will promote the developing system, produce transient science publications from ZTF alerts, and help prepare the community for LSST. 

One effort to provide observational capabilities for follow-up on the U.S.\ national facilities is the Astronomical Event Observatory Network (AEON), an initiative by NOAO that currently includes Las Cumbres, SOAR, and Gemini. The goals of this effort are to develop the necessary APIs, build on the Las Cumbres experience to provide dynamic scheduling capability for SOAR and Gemini, and coordinate data reduction and archiving efforts. So far the Las Cumbres interfaces have been updated to include the SOAR Goodman spectrograph, Goodman scripting capabilities have been implemented, and observing plans from the Las Cumbres scheduler have been executed on SOAR engineering nights. An automated, Python-based spectroscopic data reduction pipeline has been developed at SOAR that will be capable of reducing 1-D, wavelength-calibrated Goodman spectra in near real time. Starting in 2019B SOAR is planning to run TDA programs in mini-queues within observing blocks throughout the semester and is considering other options. 

Gemini will support AEON as part of the ongoing Observatory Control System (OCS) Upgrades Program and new supplemental funding from the NSF for a major project called Gemini in the Era of Multi-Messenger Astronomy (GEMMA). The Gemini observing system needs to be updated to make it easier to use, to include new features for automated scheduling, and to make it API driven. The GEMMA funds are being used for a MMA-related outreach program, a new MCAO system for Gemini North, and the real-time queue scheduler and data reduction capabilities to support AEON. The data reduction work involves transitioning the old IRAF-based code to the new DRAGONS (Python) environment and is being done in coordination with SOAR. 

AEON members and the Gemini Time Domain Astronomy Advisory Committee have been discussing new time allocation possibilities. Two options have been considered so far. In both cases the AEON members would contribute some time to an AEON pool. In option 1 a single TDA TAC would assign this time to proposals that need multiple resources on the network. This would be analogous to the current Gemini Large/Long Program TAC. In option 2, the individual AEON member TACs forward proposals that require multiple resources to a merging process, like the Gemini ITAC, that determines what projects are possible based on the TAC rankings and time available on the different facilities. 

Once Las Cumbres, SOAR, and Gemini are functioning within AEON it would be possible and very desirable to include additional facilities. A larger range of telescope apertures and geographical locations makes the network more powerful and useful for TDA/MMA follow-up. The community has a strong need for TDA capabilities on highly multiplexed spectroscopic facilities such as DESI, Subaru PFS, and the future Maunakea Spectroscopic Explorer, and on the new giant telescopes such as GMT and TMT. The AEON-compatible APIs and dynamic scheduling could be designed into their observing systems from the beginning. Finally, international facilities that can provide additional aperture and longitudinal coverage should be encouraged to join. 

\textbf{\large{5 Schedule}}

The schedules of most of the efforts described above are driven by the LSST construction and commissioning schedule. There should be a minimally functioning system with initial versions of all the components by the time that the LSST main survey is planned to begin in 2023. Continued development, and, critically, maintenance of the systems will be needed at least during the approximately 10 year time period of the LSST survey. MMA follow-up capabilities will need to be maintained for the lifetime of the survey facilities. These new systems should become common operational components of all new facilities that are involved with follow-up.

\textbf{\large{6 Cost Estimates}}

Detailed cost estimates have not been done since this is not a single project. Some representative costs based on existing efforts are:
\begin{my_itemize}
    \item Broker development: $\sim$\$16 million over 10 years
    \item TOM Toolkit development and maintenance by the Toolkit software engineers and also their curation of community contributions.: \$75,000 per year (0.5 FTE)
    \item Observation coordination system: \$500,000 over three years (3 FTE)
    \item Scheduler ``toolkit'' core development and community project coordination: \$500,000 over three years then \$75,000 per year for maintenance
    \item Development and community coordination of common data reduction tools: at least \$300,000 per year (2 FTE) 
\end{my_itemize}

Therefore, the total estimated cost over ten years is less than \$20 million, so this effort falls into the category of a small ground-based project.

{\textbf{\large{7 Summary and Recommendations}}}

Time domain and multi-messenger astrophysics are rapidly growing and important modes of observational astronomy that will help define astrophysics in the 2020s. Significant effort by a variety of groups is being put into developing the components of a follow-up system for dynamically turning survey alerts into data. These efforts need continued community support and funding in order to complete them and then maintain them for at least the lifetime of the main LSST survey. Many of the efforts such as the TOM Toolkit, scheduler toolkit, and data reduction tools, can be community open-source software projects but they will benefit from the leadership of professional software developers who can design the core infrastructure and curate the community contributions. The coordination should be done by staff of institutions that are involved in the follow-up system or MMA such as the future National Center for Observational Astronomy (NCOA, e.g.\/ LSST/Gemini/NOAO Mid-scale/Community Science and Data Center) or a new MMA institute. These tools will allow the community to produce the most science from new and exciting facilities such as LSST and improve the observing efficiency for all users of the facilities that adopt them.

\end{document}